%% file: AC_delay_isit.tex
\documentclass[10pt,conference]{IEEEtran}
\usepackage{amsmath,epsfig}

\input{new_shorts.tex}

\begin{document}

\title{Bounded Expected Delay in Arithmetic Coding}

\author{\authorblockN{Ofer Shayevitz, Ram Zamir, and Meir Feder }
\authorblockA{Tel Aviv University, Dept. of EE-Systems\\
         Tel Aviv 69978, Israel \\
         Email: \{ofersha, zamir, meir \}@eng.tau.ac.il}}

\maketitle


\begin{abstract}
We address the problem of delay in an arithmetic coding system.
Due to the nature of the arithmetic coding process, source
sequences causing {\em arbitrarily large} encoding or decoding
delays exist. This phenomena raises the question of just how large
is the expected input to output delay in these systems, i.e., once
a source sequence has been encoded, what is the expected number of
source letters that should be further encoded to allow full
decoding of that sequence. In this paper, we derive several new
upper bounds on the expected delay for a memoryless source, which
improve upon a known bound due to Gallager. The bounds provided
are uniform in the sense of being independent of the sequence's
history. In addition, we give a sufficient condition for a source
to admit a bounded expected delay, which holds for a stationary
ergodic Markov source of any order.

\end{abstract}

\section{Introduction}
Arithmetic coding has been introduced by Elias
\cite{arithmetic_coding_jelink}, as simple means to sequentially
encode a source at its entropy rate, while significantly reducing
the extensive memory usage characterizing non-sequential schemes.
The basic idea underlying this technique is the successive mapping
of growing source sequences into shrinking intervals of size equal
to the probability of the corresponding sequence, and then
representing those intervals by a binary expansion. Other coding
schemes reminiscent of Elias' arithmetic coding have been
suggested since, aimed mostly to overcome the precision problem of
the original scheme
\cite{arithmetic_coding_rissanen}\cite{arithmetic_coding_pasco}.

Delay in the classical setting of arithmetic coding stems from the
discrepancy between source intervals and binary intervals, which
may prohibit the encoder from producing bits (encoding delay) or
the decoder from reproducing source letters (decoding delay). On
top of its usual downside, delay also increases memory usage, and
therefore a large delay may turn the main advantage of arithmetic
coding on its head. As it turns out, for most sources there exists
infinite number of source sequences for which the delay is
infinite, where each sequence usually occurs with probability
zero. A well known example demonstrating this phenomena is that of
a uniform source over a ternary alphabet $\{0,1,2\}$. The source
sequence $111\ldots$ is mapped into shrinking intervals that
always contain the point $\frac{1}{2}$, and so not even a single
bit can be encoded. This observation leads to the question of just
how large is the expected delay (and consequently, the expected
memory usage) of the arithmetic coding process for a given source,
and if it is bounded at all.

The problem of delay can be practically dealt with by insertion of
a fictitious source letter into the stream to ``release'' bits
from the encoder or letters from the decoder, whenever the delay
exceed some predetermined threshold. Another possibility is coding
of finite length sequences, so that a prefix condition is
satisfied at the expense of a slightly higher redundancy, and
blocks can be concatenated \cite{IT_cover}. Nevertheless, it is
still interesting to analyze the classical sequential setting in
terms of expected delay.

In his lecture notes \cite{gallager_lecture_notes}, Gallager has
provided an upper bound for the expected delay in arithmetic
coding for a memoryless source, which was later generalized to
coding over cost channels \cite{arithmetic_coding_savari}.
Gallager's bound is given by
\begin{equation*} \Expt(D) \leq \frac{\log(8e^2\slash
\beta)}{\log(1\slash \alpha)}
\defn \mathcal{D}_g(\alpha,\beta)
\end{equation*}
where $\alpha$ and $\beta$ are the maximal and minimal source
letter probabilities respectively. Notice that this bound is
independent of the sequence's history, as shall be the case with
all the bounds presented in this paper. In Theorem
\ref{thrm:memoryless} (section \ref{sec:memoryless}) we derive a
new upper bound for the expected delay, given by
\begin{equation*}
\Expt(D) \leq 1+\frac{4\alpha\big{(}1-\alpha+ \log(1\slash
\alpha)\big{)}}{(1-\alpha)^2} \defn \mathcal{D}_1(\alpha)
\end{equation*}
which depends only on the most favorable source letter. Following
that, we show that the dependence on the least favorable letter in
Gallager's bound is unnecessary, and provide (section
\ref{sec:improving_gallager}) a uniformly tighter version of the
bound given by $\mathcal{D}_{mg}(\alpha)
\defn \mathcal{D}_g(\alpha,\alpha)$. In Theroem \ref{thrm:memoryless_tighter} (section
\ref{sec:improving_bound}) we derive another bound
$\mathcal{D}_2(\alpha)$ uniformly tighter than
$\mathcal{D}_1(\alpha)$, which is also shown to be tighter than
$\mathcal{D}_{mg}(\alpha)$ for most sources, and looser only by a
small multiplicative factor otherwise.

Our technique is extended to sources with memory, and in Theorem
\ref{thrm:memory} (section \ref{sec:memory}) we provide a new
sufficient condition for a source to have a bounded expected delay
under arithmetic coding. Specifically, this condition is shown to
hold for any stationary ergodic Markov source over a finite
alphabet.


\section{Arithmetic Coding in a Nutshell}\label{sec:perliminaries}

Consider a discrete source over a finite alphabet $\mathcal{X} =
\{0,1,\ldots,K-1\}$ with positive letter probabilities
$\{p_0,p_1,\ldots,p_{K-1}\}$. A finite source sequence is denoted
by $x_m^n = \{x_m,x_{m+1},\ldots,x_n\}$ with $x^n = x_1^n$, while
an infinite one is denoted by $x^\infty$. An arithmetic coder maps
the sequences $x^n,x^{n+1},\ldots$ into a sequence of nested
\textit{source intervals}
$\mathcal{I}(x^n)\supset\mathcal{I}(x^{n+1})\supset\ldots$ in the
unit interval that converge to a point $y(x^\infty) =
\cap_{n=1}^\infty \mathcal{I}(x^n)$ . The mapping is defined as
follows:
\begin{eqnarray*}
f_1(i) &=& \sum_{j=0}^{i-1}p_j\,,\quad f(x^1) = f_1(x_1) \\
f(x^n) &=& f(x^{n-1}) +
f_1(x_n)\Pr(x^{n-1}) \\
\mathcal{I}(x^n) &=& \left[f(x^n),f(x^n)+\Pr(x^n)\right)
\label{eq:encode}
\end{eqnarray*}
Notice that $|\mathcal{I}(x^n)| = \Pr(x^n)$ and that source
intervals corresponding to different sequences of the same length
are disjoint.
Following that, a random source sequence $X^n$ is mapped into a
random interval $\mathcal{I}(X^n)$, which as $n$ grows converges
to a random variable $Y(X^\infty)$ that is uniformly distributed
over the unit interval.

For any sequence of binary digits $b^k = \{b_1,b_2,\ldots,b_k\}$
we define a corresponding \textit{binary interval}
\begin{equation}
\mathcal{J}(b^k) = \big{[} 0.b_1b_2,\ldots b_k0,\; 0.b_1b_2,\ldots
b_k1 \big{)} \label{eq:decode}
\end{equation}
and the midpoint of $\mathcal{J}(b^k)$ is denoted by $m(b^k)$.

The process of arithmetic coding is performed as follows. The
encoder maps the input letters $x^n$ into a source interval
according to (\ref{eq:encode}), and outputs the bits representing
the smallest binary interval $\mathcal{J}(b^k)$ containing the
source interval $\mathcal{I}(x^n)$. This process is performed
sequentially so the encoder produces further bits whenever it can.
The decoder maps the received bits into a binary interval, and
outputs source letters that correspond to the minimal source
interval that contains that binary interval. Again, this process
is performed sequentially so the decoder produces further source
letters whenever it can.

\section{Memoryless Source}\label{sec:memoryless}
In this section, we provide a new bound for the expected delay of
an arithmetic coding system for a memoryless source, as a function
of the probability of the most likely source letter
\begin{equation*}
\alpha\defn\max p_k .
\end{equation*}
All logarithms in this paper are taken to the base of 2.
\begin{theorem}\label{thrm:memoryless}
\textit{Assume a sequence of $n$ source letters $x^n$ has been
encoded, and let $D$ be the number of extra letters that need to
be encoded to allow $x^n$ to be fully decoded. Then
\begin{equation}\label{eq:delay_prob}
\Pr(D > d) \leq 4\alpha^d\left (1 + d \log(1\slash \alpha)\right)
\end{equation}
independent of $x^n$. The expected delay is correspondingly
bounded by
\begin{equation}\label{eq:delay_s}
\Expt(D) \leq 1+\frac{4\alpha\big{(}1-\alpha+ \log(1\slash
\alpha)\big{)}}{(1-\alpha)^2} \defn \mathcal{D}_1(\alpha).
\vspace{0.1cm}\end{equation}}
\end{theorem}

Let us first outline the idea behind the proof. The sequence $x^n$
has been encoded into the binary sequence $b^k$ which represents
the minimal binary interval $\mathcal{J}(b^k)$ satisfying
$\mathcal{I}(x^n) \subseteq \mathcal{J}(b^k)$. The decoder has so
far been able to decode only $m<n$ letters, where $m$ is maximal
such that $\mathcal{J}(b^k) \subseteq \mathcal{I}(x^m)$. After $d$
more source letters are fed to the encoder, $x^{n+d}$ is encoded
into $b^{k'}$ where $k'\geq k$ is maximal such that
$\mathcal{I}(x^{n+d})\subseteq \mathcal{J}(b^{k'})$. Thus, the
entire sequence $x^n$ is decoded if and only if
\begin{equation}\label{eq:interval_cond}
\mathcal{I}(x^{n+d}) \subseteq \mathcal{J}(b^{k'}) \subseteq
\mathcal{I}(x^n).
\end{equation}

Now, consider the middle point $m(b^k)$, which is always contained
inside $\mathcal{I}(x^n)$ as otherwise another bit could have been
encoded. If $m(b^k)$ is contained in $\mathcal{I}(x^{n+d})$ (but
not as an edge), then condition (\ref{eq:interval_cond}) cannot be
satisfied, and the encoder cannot yield even one further bit. This
observation can be generalized to a set of points which, if
contained in $\mathcal{I}(x^{n+d})$, $x^n$ cannot be completely
{\em decoded}. For each of these points the encoder outputs a
number of bits which may enable the decoder to produce source
letters, but not enough to fully decode $x^n$. The encoding and
decoding delays are therefore treated here simultaneously, rather
than separately as in \cite{arithmetic_coding_savari}.

We now introduce some notations and prove a Lemma, required for
the proof of Theorem \ref{thrm:memoryless}. Let $[a,b)\subseteq
[0,1)$ be some interval, and $p$ some point in that interval. In
the definitions that now follow we sometime omit the dependence on
$a,b$ for brevity. We say that $p$ is {\em strictly contained} in
$[a,b)$ if $p\in [a,b)$ but $p\neq a$. We define the
\textit{left-adjacent} of $p\,$ w.r.t. $[a,b)$ to be
\begin{equation*}
\ell(p) \defn \min\left \{x\in [a,p)\,:\,\exists k\in\IntF^+ ,\, x
= p-2^{-k} \right \}
\end{equation*}
and the \textit{t-left-adjacent} of $p\;$ w.r.t. $[a,b)$ as
\begin{equation*}
\ell^{(t)}(p) \defn
\overbrace{(\ell\circ\ell\circ\cdots\circ\ell)}^t(p)\;,\quad
\ell^{(0)}(p) \defn p
\end{equation*}
Notice that $\ell^{(t)}(p) \rightarrow a$ monotonically with $t$.
We also define the \textit{right-adjacent} of $p\,$ w.r.t $[a,b)$
to be
\begin{equation*}
r(p) \defn \max\left \{x\in (p,b)\,:\,\exists k\in\IntF^+ ,\, x =
p+2^{-k} \right \}
\end{equation*}
and $r^{(t)}(p)$ as the \textit{t-right-adjacent} of $p\,$ w.r.t.
$[a,b)$ similarly, where now $r^{(t)}(p) \rightarrow b$
monotonically. For any $\delta < b-a$, the \textit{adjacent
$\delta$-set} of $p\;$ w.r.t. $[a,b)$ is defined as the set of all
adjacents that are not "too close" to the edges of $[a,b)$:
\begin{eqnarray*}
& S_\delta(p) & \defn \left \{ x\in[a+\delta,b-\delta)\,:\,
\exists\,t\in\IntF^+\cup \{0\}\,,\right. \\ & & \left.\qquad\qquad
x = \ell^{(t)}(p) \,\vee \,x = r^{(t)}(p) \right \}
\end{eqnarray*}
Notice that for $\delta > p-a$ this set may contain only
right-adjacents, for $\delta > b-p\;$ only left-adjacents, for
$\delta>\frac{b-a}{2}$ it is empty, and for $\delta=0$ it is
infinite.
\begin{lemma}\label{lem:delta_set}\textit{
The size of $S_\delta(p)$ is bounded by \vspace{-0.2cm}
\begin{equation}\label{eq_delta_set}
|S_\delta(p)|\leq 1+ 2\log{\frac{|b-a|}{\delta}}\vspace{0.2cm}
\end{equation}}
\end{lemma}
\begin{proof}
It is easy to see that the number of t-left-adjacents of $p$ that
are larger than $a+\delta$ is the number of ones in the binary
expansion of $(p-a)$ up to resolution $\delta$. Similarly, the
number of t-right-adjacents of $p$ that are smaller than
$b-\delta$ is the number of ones in the binary expansion of
$(b-p)$ up to resolution $\delta$. Defining $\lceil x \rceil^+
\defn \max (\lceil x \rceil,0 ) $, we get:
\begin{eqnarray}
\nonumber |S_\delta(p)| &\leq &
\lceil\log{\frac{p-a}{\delta}}\rceil^+ +
\lceil\log{\frac{b-p}{\delta}}\rceil^+ \\ \nonumber &\leq& \left
\{\begin{array}{ll}2+\log \frac{(p-a)(b-p)}{\delta^2} &
\,,\,\delta < p-a,b-p \\ 1+ \log{\frac{|b-a|}{\delta}} & \,,\,
o.w.
\end{array}\right.
\\ &\leq& 1+ 2\log{\frac{|b-a|}{\delta}}
\end{eqnarray}
as desired.
\end{proof}

\textit{Proof of Theorem \ref{thrm:memoryless}:} Assume the source
sequence $x^n$ has been encoded into the binary sequence $b^k$,
and let $Y = Y(x^\infty)$. Given $x^n$, $Y$ is uniformly
distributed over $\mathcal{I}(x^n)$, and thus for any interval
$\mathcal{T}$
\begin{equation}\label{eq:memoryless1}
\Pr(Y\in \mathcal{T} \,\big{|}\, x^n) = \frac{|\mathcal{T}\cap
\mathcal{I}(x^n)|}{|\mathcal{I}(x^n)|} \leq
\frac{|\mathcal{T}|}{|\mathcal{I}(x^n)|}
\end{equation}
The size of the interval $\mathcal{I}(x^{n+d})$ for $d\geq 0$ is
bounded by
\begin{eqnarray}
\nonumber\big{|}\mathcal{I}(x^{n+d})\big{|}  &=& \Pr(x^{n+d}) =
\Pr(x_{n+1}^{n+d} \big{|} x^n)\Pr({x^n}) \\ &=& \Pr(x_{n+1}^{n+d}
\big{|} x^n)|\mathcal{I}(x^n)|\leq \alpha^d
|\mathcal{I}(x^n)|\label{eq:memoryless2}
\end{eqnarray}
Combining (\ref{eq:memoryless1}) and (\ref{eq:memoryless2}), we
have that for any point $p\in \mathcal{I}(x^n)$
\begin{eqnarray}
\nonumber \Pr\Big{(}p\in \mathcal{I}(X^{n+d})\Big{|}\,x^n\Big{)}
&\leq& \Pr\Big{ (}|Y-p\,|
\leq {|}\mathcal{I}(x^n){|} \alpha^d\, \Big{|}\, x^n\Big{)} \\
&\leq& \frac{2\alpha^d|\mathcal{I}(x^n)|}{|\mathcal{I}(x^n)|} =
2\alpha^d\label{eq:memoryless3}
\end{eqnarray}
where the probabilities are taken w.r.t. to ``future'' source
letters. For any interval $\mathcal{T}\subseteq \mathcal{I}(x^n)$
that shares an edge with $\mathcal{I}(x^n)$ we have that
\begin{align}
\nonumber \Pr(\mathcal{T}\cap \mathcal{I}(X^{n+d})\neq \phi) \leq
\frac{\alpha^d|\mathcal{I}(x^n)| +
|\mathcal{T}|}{|\mathcal{I}(x^n)|} = \alpha^d +
\frac{|\mathcal{T}|}{|\mathcal{I}(x^n)|} \\ \label{eq:memoryless4}
\end{align}
For any $\delta\geq 0$, let $S_\delta$ denote the adjacent
$\delta$-set of $m(b^k)$ w.r.t. the interval $\mathcal{I}(x^n)$.
Given $x^n$, the probability that the delay $D$ is larger than $d$
is the probability that (\ref{eq:interval_cond}) is not satisfied,
which in turn is equal to the probability that the intersection
$S_0\cap\mathcal{I}(X^{n+d})$ is not empty. This fact is explained
as follows. As already shown, if $m(b^k)$ is strictly contained in
$\mathcal{I}(X^{n+d})$ then the encoder emits no further bits, and
the delay is larger than $d$. Otherwise, assume
$\mathcal{I}(X^{n+d})$ lies on the left side of $m(b^k)$.
Obviously, if $\mathcal{I}(X^{n+d})\subseteq
[\ell(m(b^k)),m(b^k))$, then $x^n$ is fully decoded since
(\ref{eq:interval_cond}) is satisfied. However, if $\ell(m(b^k))$
is strictly contained in $\mathcal{I}(X^{n+d})$ then
(\ref{eq:interval_cond}) is not satisfied, $x^n$ cannot be decoded
and the delay is larger than $d$. The same rationale also applies
to $r(m(b^k))$. Continuing the argument recursively, it is easy to
see that $x^n$ can be decoded if and only if no point of $S_0$ is
strictly contained in $\mathcal{I}(X^{n+d})$.

\begin{figure}[htbp]
  \begin{center}
    \leavevmode
    \epsfig{file=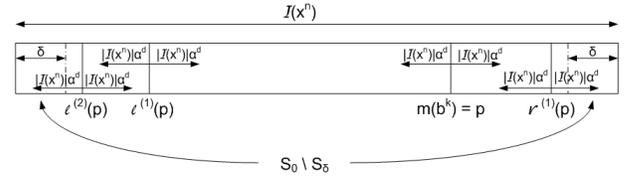, scale = 0.35}
    \caption{Source interval illustration}
    \label{fig1}
  \end{center}
  \vspace{-0.5cm}
\end{figure}

Now, notice that $S_\delta\subseteq S_0$, and that $S_0\backslash
S_\delta$ is contained in two intervals of length $\delta$ both
sharing an edge with $\mathcal{I}(x^n)$ (the situation is
illustrated in Figure \ref{fig1}). Letting $\mathcal{J}_B$ denote
a general binary interval, we bound the delay's tail probability:
\begin{eqnarray}\label{eq_delay1}
\nonumber \Pr(D > d\,\big{|}\, x^n) &=&
\Pr\left(\mathcal{I}(X^{n+d}) \not\subseteq
\mathcal{J}_B\,,\forall \mathcal{J}_B\subseteq \mathcal{I}(x^n)
\,\big{|}\,x^n\right)  \\ \nonumber &=& \Pr\left(S_0\cap
\mathcal{I}(X^{n+d}) \neq \phi\,\big{|}\,x^n\right) \leq
\\ \nonumber &\leq&  \Pr\left(S_\delta\cap \mathcal{I}(X^{n+d}) \neq
\phi\,\big{|}\,x^n\right) +\\ \nonumber &&+\;\Pr\left(\left
(S_0\backslash S_\delta\right)\cap \mathcal{I}(X^{n+d}) \neq
\phi\,\big{|}\,x^n\right) \\ \nonumber &\leq& 2\alpha^d|S_\delta
| + 2\left(\alpha^d + \frac{\delta}{|\mathcal{I}(x^n)|}\right) \leq \\
\nonumber &\leq&
\;2\alpha^d\left(1+2\log{\frac{|\mathcal{I}(x^n)|}{\delta}}\right)+
\\ &&+ 2\left(\alpha^d + \frac{\delta}{|\mathcal{I}(x^n)|}\right)
\end{eqnarray}
Lemma \ref{lem:delta_set} and equations
(\ref{eq:memoryless3}),(\ref{eq:memoryless4}) were used in the
transitions. Taking the derivative of the right-hand-side of
(\ref{eq_delay1}) w.r.t. $\delta$ we find that $\delta = 2\alpha^d
|\mathcal{I}(x^n)|$ minimizes the bound. We get:
\begin{eqnarray*}
\Pr(D > d\,\big{|}\, x^n) &\leq&
2\alpha^d\left(1+2\log{\frac{1}{2\alpha^d}}\right ) + 6\alpha^d
\\ &=& 4\alpha^d\left (1 + d \log(1\slash \alpha)\right)
\end{eqnarray*}
and (\ref{eq:delay_prob}) is proved. Now, the expectancy of $D$
given $x^n$ can be bounded accordingly
\begin{eqnarray}\label{eq:memoryless5}
\nonumber \Expt \,(D\big{|}\,x^n) &=& \sum_{d=1}^\infty
d\Pr(D=d\,\big{|}\,x^n) = \sum_{d=1}^\infty \Pr(D\geq d\,\big{|}\,
x^n) \\ \nonumber &\leq& 1+\sum_{d=1}^\infty \Pr(D > d\,\big{|}\,
x^n) \\ \nonumber &\leq& 1+4\sum_{d=1}^\infty \alpha^d\left (1 + d
\log(1\slash \alpha)\right) \\
&=& 1+\frac{4\alpha\big{(}1-\alpha+ \log(1\slash
\alpha)\big{)}}{(1-\alpha)^2}
\end{eqnarray}
and (\ref{eq:delay_s}) is proved. Notice that both of the bounds
above are uniform so the dependence on $x^n$ can be removed.
\endproof

\section{Improving Gallager's Bound}\label{sec:improving_gallager}
Gallager \cite{gallager_lecture_notes} provided an upper bound for
the expected delay in arithmetic coding of a memoryless source,
given by
\begin{equation*}
\Expt(D) \leq \frac{\log(8e^2\slash \beta)}{\log(1\slash \alpha)}
\defn \mathcal{D}_g(\alpha,\beta)
\end{equation*}
where $\alpha \defn \max p_k$ and $\beta \defn \min p_k$. Notice
that our bound $\mathcal{D}_1(\alpha)$ in (\ref{eq:delay_s})
depends only on the most likely source letter, while Gallager's
bound $\mathcal{D}_g(\alpha,\beta)$ depends also on the least
likely source letter. Moreover, holding $\alpha$ constant we find
that $\mathcal{D}_g(\alpha,\beta)\tendsto{\beta}{0}\infty$. This
phenomena is demonstrated in the following example.

\begin{figure}[htbp]
  \begin{center}
    \leavevmode
    \epsfig{file=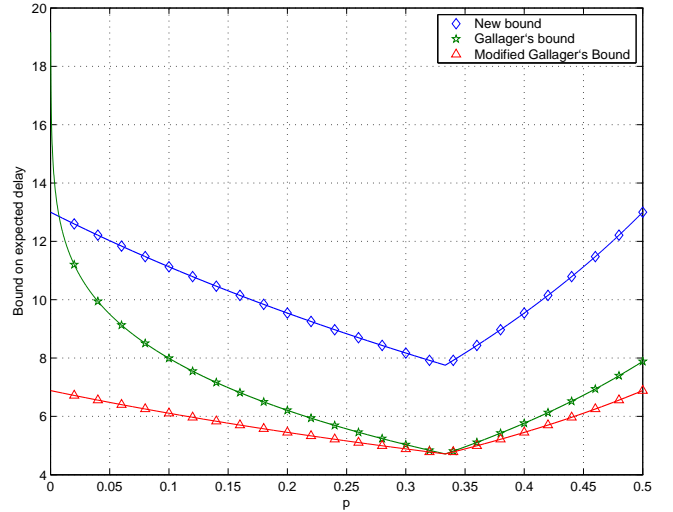, scale = 0.5}
    \caption{Bounds for the ternary source}
    \label{ternary2}
  \end{center}
  \vspace{-0.5cm}
\end{figure}

\textit{Example: Consider a ternary source with letter
probabilities $(p,\frac{1-p}{2},\frac{1-p}{2})$. Both bounds for
that source are depicted in Figure \ref{ternary2} as a function of
$p$, together with a modified bound derived in the sequel. As can
be seen, Gallager's bound is better for most values of $p$, but
becomes worse for small $p$, due to its dependence on the least
probable source letter. In fact, the bound {\em diverges} when
$p\rightarrow 0$, which is counterintuitive since we expect the
delay in this case will approach that of a uniform binary source
(for which $\mathcal{D}_g(\alpha,\beta)$ is finite). In contrast,
the new bound which depends only on the most likely letter tends
to a constant when $p\rightarrow 0$, which equals its value for
the corresponding binary case.}

Intuition suggests that least likely letters are those that tend
to accelerate the coding/decoding process, and that the dominating
factor influencing the delay should be the most likely source
letters. Motivated by that, we turn to examine the origin of the
term $\beta$ in Gallager's derivations.



Gallager's bound for the expected delay is derived via a
corresponding bound on the {\em information delay}, i.e., the
difference in self-information between a source sequence and an
extended source sequence needed to ensure that the original
sequence is completely decoded. We remind the reader that the self
information of a sequence $x^n$ is just $-\log{(\Pr(x^n))}$. We
now follow the derivations in \cite{arithmetic_coding_savari},
replacing notations with our own and modifying the proof to remove
the dependence on $\beta$. Notice that
\cite{arithmetic_coding_savari} analyzes the more general setting
of cost channels which reduces to that of
\cite{gallager_lecture_notes} and to ours and by setting $ N=2,
C=c_i=c_{max}=1$ (in the notation therein).

Consider a source sequence encoded by a binary sequence $b^k$. A
bound on the expected self-information of that sequence with the
last letter truncated is given by \cite[equations
10,11]{arithmetic_coding_savari}
\begin{equation}\label{eq:gallager1}
\Expt \big{(}I(x^{n(k)-1})\big{|} b^k\big{)} \leq k + \log (2e)
\end{equation}
where $n(k)$ is the number of source letters emitted by the
source, and $I(x^{n(k)-1})$ is the self-information of the
corresponding source sequence without the last letter. Using the
relation $I(x^n) \leq I(x^{n-1}) + \log (1\slash\beta)$, we get a
bound on the self-information of the sequence \cite[equation
14]{arithmetic_coding_savari}:
\begin{equation}\label{eq:gallager2}
\Expt \big{(}I(x^{n(k)})\big{|} b^k\big{)} \leq k + \log (2e\slash
\beta)
\end{equation}
This is the only origin of the term $\beta$. In order to obtain a
bound on the expected information delay, there seems to be no
escape from the dependence on $\beta$. However, we are interested
in the delay in source letters. We therefore continue to follow
\cite{arithmetic_coding_savari} but use (\ref{eq:gallager1}) in
lieu of (\ref{eq:gallager2}) to bound the information delay up to
one letter before the last needed for decoding. This approach
eliminates the dependence on the least likely letter, which if
appears last may increase the self-information considerably but
meanwhile contribute only a single time unit to the delay.

Consider a specific source sequence $x^n$. A bound on the expected
number of bits $k(n)$ required to decode that sequence is given by
\cite[equation 15]{arithmetic_coding_savari}:
\begin{equation}\label{eq:gallager3}
\Expt \big{(}k(n)\big{|}x^n\big{)} \leq I(x^n) + \log (4e)
\end{equation}
Now, let $b^{k(n)}$ be the binary sequence required to decode
$x^n$. Using (\ref{eq:gallager1}) (instead of (\ref{eq:gallager2})
used in \cite{gallager_lecture_notes} and
\cite{arithmetic_coding_savari}) we have that
\begin{equation}\label{eq15}
\Expt \big{(}I(x^{n+D-1})\big{|}b^{k(n)},x^n\big{)} \leq k(n) +
\log (2e)
\end{equation}
where $D$ is the number of extra letters needed to ensure the
encoder emits the necessary $k(n)$ bits. Using
(\ref{eq:gallager3}) and taking the expectation w.r.t. $k(n)$ we
find that
\begin{equation}\label{eq:gallager4}
\Expt \big{(}I(x^{n+D-1})\big{|}x^n\big{)} -I(x^n) \leq \log
(8e^2)
\end{equation}
and the modified bound for the delay in source letters follows
through by dividing (\ref{eq:gallager4}) by the minimal letter
self-information $\log (1\slash \alpha)$ and rearranging the
terms:
\begin{equation}\label{eq:gallager5}
\Expt \big{(}D\big{|}x^n\big{)} \leq 1+\frac{\log (8e^2)}{\log
(1\slash \alpha)} \defn \mathcal{D}_{mg}(\alpha)
\end{equation}
Notice that the modified Gallager bound
$\mathcal{D}_{mg}(\alpha)=\mathcal{D}_g(\alpha,\alpha)$ is
uniformly lower $\mathcal{D}_g(\alpha,\beta)$, and coincides with
it only for uniformly distributed sources.

\textit{Example (continued): The modified Gallager bound for the
ternary source converges for $p\rightarrow 0$, as illustrated in
Figure \ref{ternary2}. It is also easy to verify that it converges
to the same value it takes for a uniform binary source.}

%

The ratio of our bound to the modified Gallager bound is depicted
in Figure \ref{fig:ratios}, together with two tighter bounds
introduced in the following section. Comparing the bounds, we find
that $\mathcal{D}_1(\alpha)$ is at most $\sim 2.4$ times worse
than $\mathcal{D}_{mg}(\alpha)$, and is even better for small
(below $\sim 0.069$) values of $\alpha$. For $\alpha\rightarrow 0$
the ratio tends to unity, since both $\mathcal{D}_1(\alpha)$ and
$\mathcal{D}_{mg}(\alpha)$ approach 1, the minimal possible delay
for a source that is not 2-adic. Indeed, for a very small $\alpha$
it is intuitively clear that even when a single extra letter is
encoded, the source interval decreases significantly which enables
decoding of the preceding source interval with high probability.

\section{Improving Our Bound}\label{sec:improving_bound}
As we have seen, $\mathcal{D}_1(\alpha)$ is good for small values
of $\alpha$ (the probability of the most likely letter) and
becomes worse for larger values. The source of this behavior lies
in a somewhat loose analysis of the size of $S_\delta$ for large
$\delta$, and also since for large $\alpha$ and small $d$ the
bound (\ref{eq:delay_prob}) may exceed unity. A more subtle
analysis enables us to improve our bound for large $\alpha$, and
the result is now stated without proof.
\begin{theorem}\label{thrm:memoryless_tighter}
\textit{Let $d_0 = \left\lfloor
\frac{2}{\log(1\slash\alpha)}\right\rfloor$, and define $d_1\geq
d_0$ to be the largest such integer for which every integer $d_0 <
d\leq d_1$ (if there are any) satisfies
\begin{equation*}
2\alpha^d\left (1 + 2d \log(1\slash \alpha)\right) > 1
\end{equation*}
The expected delay of an arithmetic coding system for a memoryless
source is bounded by
\begin{eqnarray}
&&\Expt \,(D) \leq  \mathcal{D}_2(\alpha) \defn \\ &&\nonumber 1+
d_1 + \frac{2\alpha^{d_1+1}}{1-\alpha} +
\frac{4\alpha^{d_1+1}\big{(}d_1(1-\alpha)+1\big{)}\log(1\slash
\alpha)}{(1-\alpha)^2}
\end{eqnarray}}

\end{theorem}

\begin{figure}[htbp]
  \begin{center}
    \leavevmode
    \epsfig{file=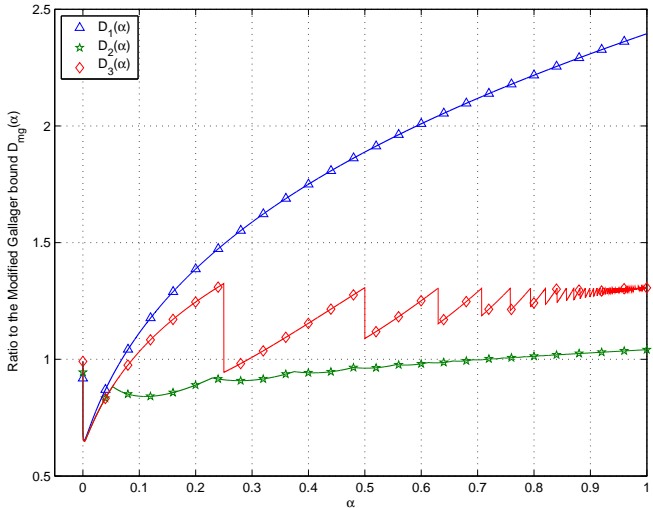, scale = 0.5}
    \caption{The ratio of the different bounds to the modified Gallager bound}
    \label{fig:ratios}
  \end{center}
    \vspace{-0.2cm}
\end{figure}

An explicit bound $\mathcal{D}_3(\alpha)$ (though looser for large
$\alpha$) can be obtained by substituting $d_1 = d_0$. The ratio
of our original bound $\mathcal{D}_1(\alpha)$, the modified bound
$\mathcal{D}_2(\alpha)$ and its looser version
$\mathcal{D}_3(\alpha)$ to the modified Gallager bound
$\mathcal{D}_{mg}(\alpha)$ are depicted in Figure
\ref{fig:ratios}. As can be seen, $\mathcal{D}_2(\alpha)$ is
tighter than $\mathcal{D}_{mg}(\alpha)$ for values of $\alpha$
smaller than $\sim 0.71$, and for larger values is looser but only
up to a multiplicative factor of $\sim 1.04$. Notice again that
all of the bounds coincide for $\alpha\rightarrow 0$, as in this
case they all tend to 1 which is the best possible general upper
bound.

\section{Sources with Memory}\label{sec:memory}
The discussion of section \ref{sec:memoryless} is easily
generalized to sources with memory. The only point in the proof
that needs to be reestablished is the definition of $\alpha$,
which was the probability of the most likely source letter in the
memoryless case.
\begin{theorem}\label{thrm:memory}
\textit{Consider an arithmetic coding system for a source with a
probability distribution $p(x^n)$ over a finite alphabet
$\mathcal{X}$. Let
\begin{equation*}
\gamma(d) \defn \sup_n
\max_{x^{n+d}\in\mathcal{X}^{n+d}}p(x^{n+d}\,\big{|}\,x^n)
\end{equation*}
If $\,\gamma(d) = o\left(d^{\,-(1+\varepsilon)}\right)$ for some
$\varepsilon>0$, then the expected delay of the system is
bounded.}
\end{theorem}
\begin{proof}
The derivations for the memoryless case can be repeated, with
$\alpha^d$ replaced by $\gamma(d)$. The bound
(\ref{eq:memoryless5}) becomes
\begin{equation*} \Expt \,(D\,\big{|}\,x^n) \leq 1+
4\sum_{d=1}^\infty \gamma(d)\left (1 +
\log{\frac{1}{\gamma(d)}}\right)
\end{equation*}
If the sum above converges, then we have the bounded expected
delay property. The condition given in Theorem \ref{thrm:memory}
is sufficient to that end.
\end{proof}

For a memoryless source, $\gamma(d) = \alpha^d$ and the condition
is satisfied. It is also fulfilled for any source with memory
whose conditional letter probabilities are bounded away from 1,
and thus such sources admit a bounded expected delay. This fact
was already observed in \cite{arithmetic_coding_savari} with the
additional requirement for the conditional probabilities to be
bounded away from 0 as well (a byproduct of the dependency on the
least favorable letter). The condition in Theorem
\ref{thrm:memory} is however more general. As an example, consider
a stationary ergodic first order Markov source. Such a source
satisfies
\begin{equation}\label{eq:memory_p1}
p(x^{n+|\mathcal{X}|}\,\big{|}\,x^n) < 1\,,\quad
\forall\,x^{n+|\mathcal{X}|}\in \mathcal{X}^{n+|\mathcal{X}|}
\end{equation}
since otherwise the source would have a deterministic cycle which
contradicts the ergodic assumption. Define:
\begin{equation*}
\xi \defn \max_{x^{n+|\mathcal{X}|}\in
\mathcal{X}^{n+|\mathcal{X}|}}
p(x^{n+|\mathcal{X}|}\,\big{|}\,x^n)
\end{equation*}
We have from (\ref{eq:memory_p1}) that $\xi < 1$, and since the
source is stationary, $\xi$ is also independent of $n$.
$\gamma(d)$ is monotonically non-increasing and therefore
$\gamma(d) \leq
\xi^{\left\lfloor\frac{d}{|\mathcal{X}|}\right\rfloor}$ and is
exponentially decreasing with $d$, thus satisfying the condition
in Theorem \ref{thrm:memory}. This result can be generalized to
any Markov order.
\begin{corollary}
\textit{The expected delay of arithmetic coding for a finite
alphabet, stationary ergodic Markov source of any order is
bounded.}
\end{corollary}

\section{Summary}\label{sec:summary}
New upper bounds on the expected delay of an arithmetic coding
system for a memoryless source were derived, as a function of the
probability $\alpha$ of the most likely source letter. In
addition, a known bound due to Gallager that depends also on the
probability of the least likely source letter was uniformly
improved by disposing of the latter dependence. Our best bound was
compared to the modified Gallager bound, and shown to be tighter
for $\alpha < 0.71$ and looser by a multiplicative factor no
larger than $\sim 1.04$ otherwise. The bounding technique was
generalized to sources with memory, providing a sufficient
condition for a bounded delay. Using that condition, it was shown
that the bounded delay property holds for a stationary ergodic
Markov source of any order.

Future research calls for a more precise characterization of the
expected delay in terms of the entire probability distribution,
which might be obtained by further refining the bounding technique
presented in this paper. In addition, a generalization to coding
over cost channels and finite-state noiseless channels in the
spirit of \cite{arithmetic_coding_savari} can be considered as
well.

\bibliographystyle{IEEEbib}
\bibliography{ofer_refs}

%
%
%
%
%

\end{document}

%% file: new_shorts.tex
\newtheorem{lemma}{Lemma}
\newtheorem{corollary}{Corollary}
\newtheorem{theorem}{Theorem}

\newcommand{\bre}{\begin{equation}}
\newcommand{\ere}{\end{equation}}
\newcommand{\be}\[
\newcommand{\ee}\]
\newcommand{\bra}{\begin{eqnarray}}
\newcommand{\era}{\end{eqnarray}}
\newcommand{\ba}{\begin{eqnarray*}}
\newcommand{\ea}{\end{eqnarray*}}
\newcommand{\bfg}{\begin{figure}[hbtp]}
\newcommand{\efg}{\end{figure}}
\newcommand{\bver}{\begin{verbatim}}
\newcommand{\ever}{\end{verbatim}}
\newcommand{\bit}{\begin{itemize}}
\newcommand{\eit}{\end{itemize}}
\newcommand{\ben}{\begin{enumerate}}
\newcommand{\een}{\end{enumerate}}

\newcommand{\comment}[1]{}

\newlength{\tmpbigbar}

\newcommand{\tendsto}[2]
{\raisebox{-1.2ex}{$\stackrel{\textstyle \longrightarrow}{
\scriptscriptstyle #1\rightarrow #2}$}}

\newfont{\boldlarge}{msbm10 scaled 1100}

\newcommand{\Expt}{\mbox{\boldlarge E}}
\newcommand{\defn}{ \stackrel{\triangle}{=} }
\newcommand{\IntF}{\mbox{\boldlarge Z}}